\documentclass[conference,a4paper]{APSIPA2021}
\usepackage{amsmath}
\usepackage{graphicx}
\usepackage{multirow}
\usepackage{threeparttable}
\usepackage[backend=biber,style=ieee,]{biblatex}

\usepackage{geometry}
\usepackage{fancyhdr}

\fancypagestyle{firststyle}{
  \fancyhf{}
  \fancyhead[C]{2024 Asia Pacific Signal and Information Processing Association Annual Summit and Conference (APSIPA ASC)}
}

\begin{document}

\title{Frequency \& Channel Attention Network for Small Footprint Noisy Spoken Keyword Spotting}

\author{
\authorblockN{
Yuanxi Lin and
Yuriy Evgenyevich Gapanyuk
}

\authorblockA{
Bauman Moscow State Technical University, Russia \\
linyu@student.bmstu.ru, gapyu@bmstu.ru}

}

\maketitle
\thispagestyle{firststyle}
\pagestyle{fancy}

\begin{abstract}
In this paper, we aim to improve the robustness of Keyword Spotting (KWS) systems in noisy environments while keeping a small memory footprint. We propose a new convolutional neural network (CNN) called FCA-Net, which combines mixer unit-based feature interaction with a two-dimensional convolution-based attention module. First, we introduce and compare lightweight attention methods to enhance noise robustness in CNN. Then, we propose an attention module that creates fine-grained attention weights to capture channel and frequency-specific information, boosting the model's ability to handle noisy conditions. By combining the mixer unit-based feature interaction with the attention module, we enhance performance. Additionally, we use a curriculum-based multi-condition training strategy. Our experiments show that our system outperforms current state-of-the-art solutions for small-footprint KWS in noisy environments, making it reliable for real-world use.
\end{abstract}

\section{Introduction}

Keyword Spotting (KWS)~\cite{9665775} is a technology that detects specific keywords in audio streams. It aims to recognize trigger words or commands, allowing smart devices to work without physical contact. This technology is widely used in consumer electronics such as smart speakers, wearables, voice assistants, and Internet of Things devices~\cite{9099251}. We focus on small-footprint KWS in this paper, which requires a small memory footprint and low computational power. This kind of technology is practical because the KWS models are always powered on low-resource devices to provide a constant service.

Recent small-footprint KWS models based on the deep neural network (DNN) have achieved success with less noisy, close-talking audio datasets. Chen et.al~\cite{chen2014small} first utilize the DNN to regard keyword spotting as a classification problem. 
Tang and Lin~\cite{tang2018deep} proposed a residual network (ResNet) based KWS system where they used dilated convolution to enlarge the size of the receptive field exponentially with the depth of the network. Choi et.al~\cite{choi19_interspeech} propose a temporal convolutional neural network (TC-ResNet) for real-time KWS on mobile devices, with temporal convolution reducing computation and increasing performance in terms of accuracy compared to 2D convolutions. 
However, these systems face significant challenges in noisy environments because background noise can interfere with the detection of keywords, leading to false activations or missed detections. 

Improving noise robustness in KWS systems is a crucial task to ensure reliable performance~\cite{lin2024advancing}. The motivation to handle this kind of environment is driven by the need for KWS to function accurately in real-world settings where noise is often present, such as in busy households, public spaces, and industrial areas. Enhancing noise robustness will make KWS systems more effective and user-friendly, ensuring they can be relied upon in diverse and challenging auditory conditions.


To augment the noise robustness of acoustic models, curriculum learning has been proposed~\cite{braun2017curriculum}. 
As a learning method, curriculum learning increases the difficulty of training samples progressively. It starts with initial training on easy, clean samples and then continues with difficult samples containing loud noise, similar to how a person studies according to a curriculum. Curriculum learning is widely used for noise-robust KWS~\cite{higuchi2021dynamic}.


Recent work on Ng et.al~\cite{ng2022convmixer,xiao3} for small-footprint KWS addresses the challenge of noisy environments. They propose a new encoder architecture based on a mixer module. This mixer computes weighted feature interactions across global channels, improving information flow. The model ConvMixer shows good performance in noisy, far-field conditions. However, there is still room to improve with efficient attention modules to extract local features, such as SE~\cite{hu2018squeeze} and ECA~\cite{wang2020eca}. This improvement is important because local features can enhance the model's ability to handle varying noise levels and complex audio inputs, leading to more accurate keyword spotting.


In this paper, we aim to enhance the robustness of KWS in noisy environments while ensuring a small memory footprint. We propose a novel network called FCA-Net combining feature interaction in ConvMixer with an efficient attention module. First, we introduce and compare various lightweight attention methods to improve the noise robustness of KWS systems. Then, we propose an efficient two-dimensional convolution-based attention module for noisy KWS. This module produces fine-grained attention weights to capture channel and frequency-specific information, enhancing the model's representation ability for noise-robust speech. Additionally, we implement a curriculum-based multi-condition training strategy. Our experiments show that our system outperforms existing state-of-the-art solutions for small-footprint KWS in noisy conditions.


\section{Related work}

{\bf Small Footprint Keyword Spotting } - With the widespread adoption of voice interfaces in smart consumer electronics, the application of small convolutional neural networks (CNNs) in compact keyword spotting has become increasingly significant. Recent works investigated innovative convolution techniques to improve KWS performance. TC-ResNet~\cite{choi19_interspeech,xiao2} applies 1D temporal convolution to enhance efficiency and accuracy. BC-ResNet~\cite{kim21l_interspeech} introduces broadcasted residual learning combining 1D and 2D convolutions. 
However, most studies do not consider noise robustness, which is a crucial factor for successful voice interfaces.

{\bf Noise robust Spoken Keyword Spotting} - Data augmentation, especially noise mixture training, has been widely used to enhance the noise robustness of acoustic models~\cite{trinh2022importantaug}. Noise mixture training involves adding environmental noise to the training data, enabling the model to better handle noisy environments. However, small neural networks struggle to learn in high-noise conditions. To address this issue, curriculum learning has been introduced, which gradually increases the difficulty of training samples to improve the model's robustness~\cite{braun2017curriculum}. Although these methods have improved performance to some extent, they still fall short when dealing with high-noise environments.

{\bf Attention mechanisms in CNNs} -  Attention mechanism address the limitations of traditional CNNs in handling long-range dependencies and feature selection. It aims to enable the model to automatically focus on more important information regions. SE-Net~\cite{hu2018squeeze,fmsg2} introduces an  effective mechanism for learning channel attention, achieving significant and promising performance outcomes for the first time. Subsequently, Woo et al.~\cite{woo2018cbam} proposed the Convolutional Block Attention Module (CBAM), which further enhanced model performance by combining channel attention and spatial attention. Attention mechanisms, while powerful, often come with significant computational and memory overheads. These overheads can hinder the deployment of models on edge devices. Consequently, recent research has focused on developing light attention mechanisms. ECA-Net~\cite{wang2020eca} simplifies the attention mechanism by using an efficient channel attention module that avoids dimensionality reduction and performs local cross-channel interaction without significantly increasing the computational complexity. However, integrating these attention mechanisms with the noise-robust spoken keyword spotting systems is still under exploration. 
 
\begin{figure}[htbp]
\begin{center}
\includegraphics[width=\linewidth]{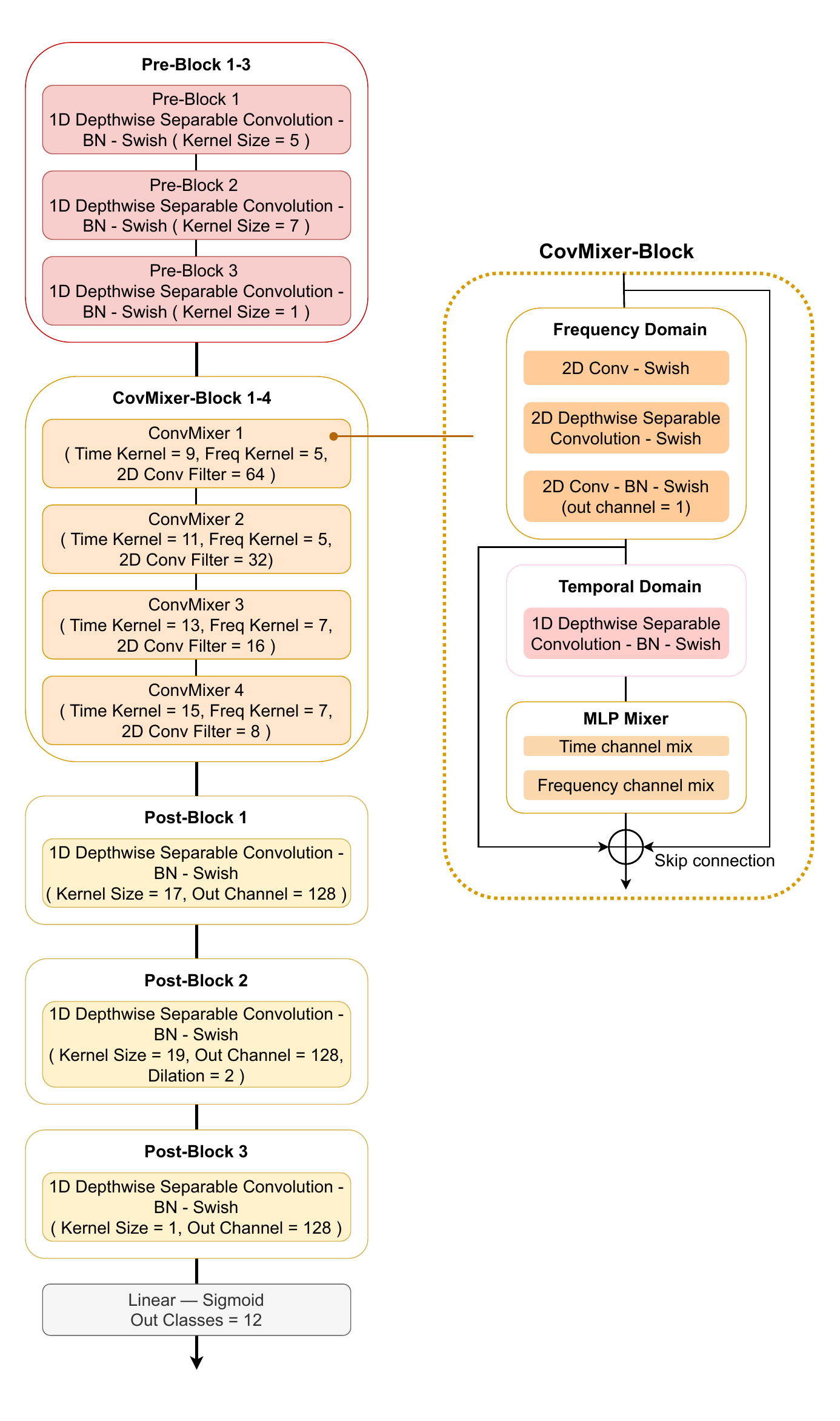}
\label{convmixer}
\end{center}
\caption{Overview of ConvMixer model architecture.}
\vspace*{-20pt}
\end{figure}

\section{Methodology}
To overcome the difficulty of learning the noise-robustness property in small-footprint keyword spotting, we propose a joint framework that combines ConvMixer block and attention mechanism.

\subsection{ConvMixer architecture}

\begin{figure*}[htbp]
\centering
\begin{minipage}[t]{0.48\linewidth}
\includegraphics[width=\linewidth]{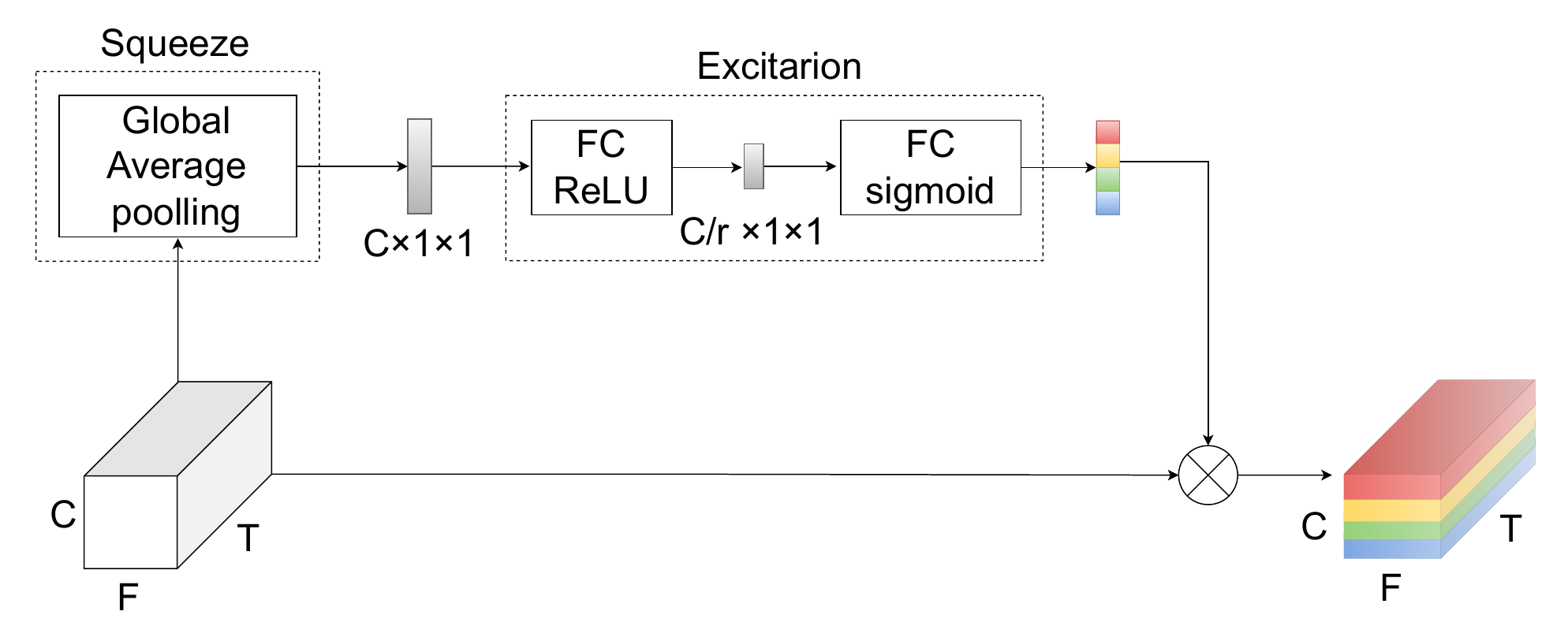}
\centerline{(a)}
\end{minipage}
\hfill
\begin{minipage}[t]{0.48\linewidth}
  \centering
  \includegraphics[width=\linewidth]{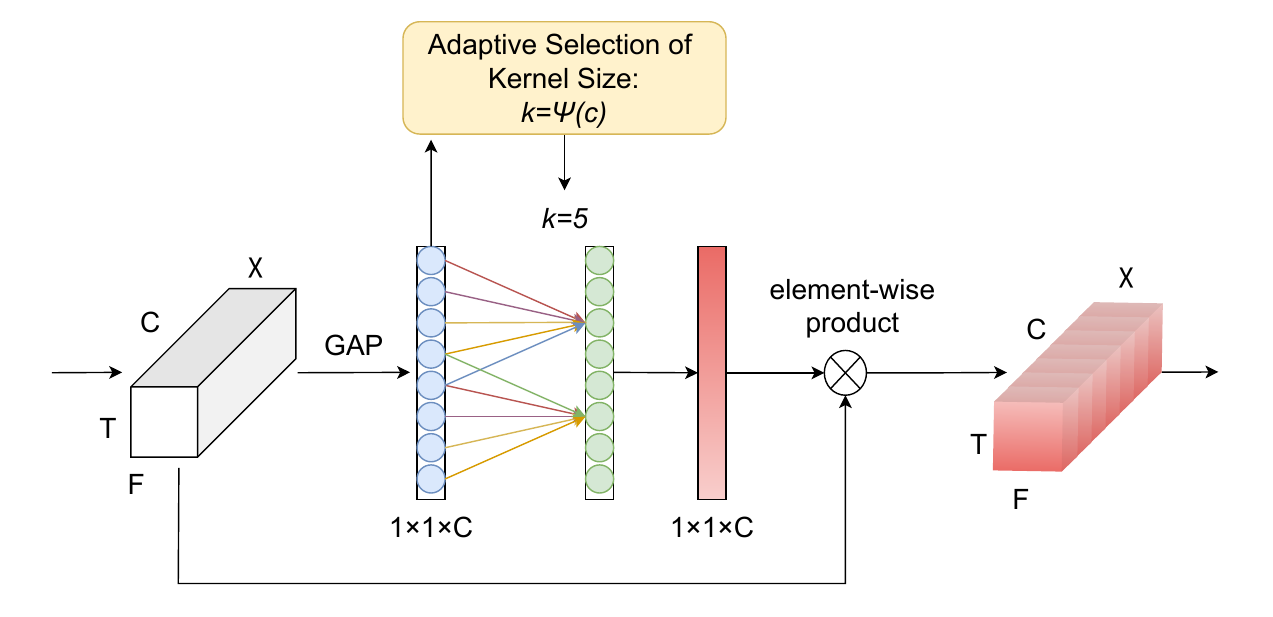}
  \centerline{(b)}
  \end{minipage}
\hfill
\caption{An illustration of SE, ECA attention. GAP means global average pooling.}
\end{figure*}

The ConvMixer model architecture is divided into three main sections: the pre-convolutional block, the ConvMixer block, and the post-convolutional block. Figure \ref{convmixer} provides an overview of this architecture.

The pre-convolutional and post-convolutional block includes 1D depthwise separable (DWS) convolution, batch normalization, and swish activation.

The ConvMixer block integrates 2D and 1D DWS convolutions to extract features from both the frequency and temporal domains. It also incorporates a mixer layer to enhance information flow. The following equations describe the ConvMixer block:
\vspace{-5pt}
\begin{equation}
z = \sigma \circ f_1(\sigma \circ f(x))
\end{equation}

 In the ConvMixer block, \( f \) is a 2D-DWS convolution, and \( \sigma \) is the Swish activation function. First, the input \( x \) is convolved by the 2D-DWS convolution function \( f \), and then the result is processed by the activation function \( \sigma \) to obtain the intermediate features. These intermediate features are convolved again by the 2D-DWS convolution \( f \), and processed by the activation function \( \sigma \), resulting in the frequency domain features \( z \).

\begin{equation}
y_1 = \sigma \circ \text{BatchNorm}(f(z))
\end{equation}

Then the frequency domain features \( z \) is convolved by a 1D-DWS function \( f_2 \). The result is processed by a batch normalization (BatchNorm) layer and finally activated by the function \( \sigma \), resulting in the temporal domain features \( y_1 \).

\begin{equation}
y_2 = \sigma \circ \text{BatchNorm}(f_2(y_1))
\end{equation}

The temporal domain features \( y_1 \) are processed by another 1D-DWS convolution function \( f_2 \), then passed through a BatchNorm layer and activation, resulting in the processed features \( y_2 \).

\begin{equation}
\tilde{y} = x + y_1 + f_3(y_2)
\end{equation}

Finally, the input \( x \), intermediate features \( y_1 \), and processed features \( y_2 \) are added together to obtain the output \( \tilde{y} \) of the ConvMixer module. Eq (4) computes the processed features \( y_2 \) of the block with \(f_3\) as the mixer layer.  We propose using two types of multi-layer perceptrons (MLP): temporal channel mixing and frequency channel mixing, to induce interaction between the feature space. Each MLP mixing involves two linear layers and a GELU activation unit, independently processing each temporal and frequency channel.

\subsection{Efficient attention module}
\subsubsection{Squeeze-and-excitation block (SE)}
The Squeeze-and-Excitation (SE) block~\cite{hu2018squeeze} is an efficient attention mechanism used in CNNs to improve performance. As shown in Figure 2 (a), the SE block has two main operations: Squeeze and Excitation. The Squeeze operation performs global average pooling on the feature maps across the channel dimension to create a compact intermediate representation. The Excitation operation uses two fully connected (FC) layers to assign importance weights to each channel. The SE module re-weights the output of each channel, highlighting important features and suppressing less significant ones. This helps the network focus on the most informative features, enhancing performance with minimal additional computational cost.

\subsubsection{Efficient channel attention (ECA)}

The Efficient Channel Attention (ECA)~\cite{wang2020eca} block is a highly efficient channel attention mechanism for deep CNNs. As illustrated in Figure 3, the ECA block first uses global average pooling to aggregate features. It then applies a 1D convolution with an adaptively determined kernel size, generating channel weights through a Sigmoid function. Unlike other methods, the ECA block avoids dimensionality reduction via fully connected layers, maintaining a direct correspondence between channels and their weights.

The ECA block captures local cross-channel interactions through 1D convolution, improving performance while maintaining efficiency. The kernel size 
\(k\) of this 1D convolution is adaptively determined based on the channel dimension \(C\) , ensuring that higher-dimensional channels interact more broadly, while lower-dimensional channels interact more narrowly. The ECA block achieves effective channel attention with low model complexity, significantly enhancing performance and demonstrating strong generalization in practical tasks.


\subsubsection{Convolution-based two-dimensional attention (C2D)}

The C2D block is integrated into CNNs to calculate fine-grained attention weights for various channel and frequency positions. This module enhances the efficiency and effectiveness of the attention mechanism by using convolution operations instead of the fully connected layers typically used in traditional attention mechanisms which shown in Figure 4.


The input feature map $X$ of C2D block contains $C$ channels, $F$ frequency bins, and $T$ time frames. This feature map is the output of a previous convolutional layer in the network. Then the Global Average Pooling (GAP) is applied along the time dimension of the input feature map $X$.

 \begin{equation}
z(c, f) = \frac{1}{T} \sum_{t=1}^{T} X(c, f, t)
\end{equation}

The result is a channel-frequency plane $z$ with dimensions $C \times F$, where each element represents the average value of the corresponding channel and frequency bin across the time dimension. Next, the pooled channel-frequency plane $z$ is passed through two consecutive 2D convolutional layers. These convolution operations are used to calculate the attention weights $\omega$ for different channel and frequency positions. The convolutional layers include BatchNorm and ReLU activation functions to enhance learning ability and stability. The attention weights $\omega$ represent the importance of each channel-frequency pair.

\begin{equation}
\omega = \sigma(\text{Conv2}(\text{ReLU}(\text{BN}(\text{Conv1}(z)))))
\end{equation}

The calculated attention weights $\omega$ are then applied to the original input feature map $X$ through element-wise multiplication. This step re-scales the input features, enhancing important features and suppressing less important ones as the following equation:


\begin{equation}
X' = \omega \odot X
\end{equation}
    
This operation re-scales the input features, emphasizing important channel-frequency pairs and suppressing less important ones, thereby enhancing the representation capability of the network.

The C2D block calculates attention weights using convolution operations, which require fewer parameters and less computational cost compared to fully connected layers. By simultaneously considering both channel and frequency dimensions, the C2D block generates more detailed and specific attention weights, enhancing feature representation capability.

\subsection{Proposed FCA-Net}
Our proposed FCA-Net combines the ConvMixer with the C2D attention mechanism for keyword spotting. We add the C2D attention after each ConvMixer block, referring to this configuration as FCA-Net-all. Additionally, we explore other variants: FCA-Net-pre, which adds C2D attention only to the pre-convolutional blocks; FCA-Net-post, which adds C2D attention only to the post-convolutional blocks; and FCA-Net-final, which adds C2D attention before the final linear layer to study its impact on the final classification. We setup various experiments for different variants in the ablation study.
\begin{figure}[tbp]
\centering
\includegraphics[width=\linewidth]{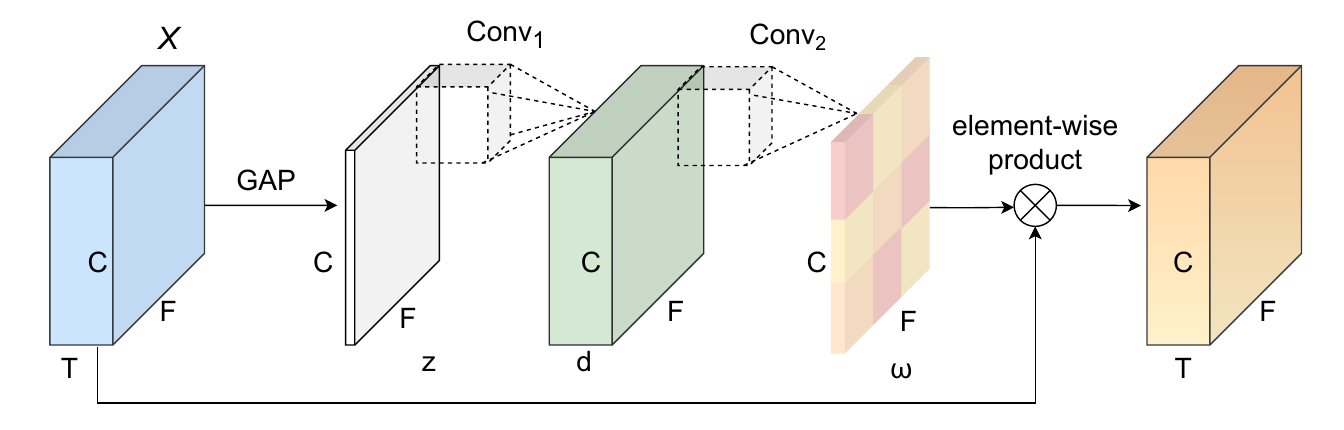}
\caption{The network structure of the C2D block.}
\end{figure}

\begin{table*}[ht]
    \centering
        \caption{Comparison with the SOTA models.}
    \begin{tabular}{c|c|c|c|c|c|c|c|c}
        \hline
        \multirow{2}{*}{\textbf{Model}} & \multirow{2}{*}{\textbf{Num. of Params (K)}} & \multirow{2}{*}{\textbf{MACs (M)}} & \multirow{2}{*}{\textbf{Acc. of V2-12 Official (\%)}} & \multicolumn{5}{c}{\textbf{Accuracy of Noisy Test Command(\%)}} \\
        \cline{5-9}
        & & & & \textbf{Clean} & \textbf{20 dB} & \textbf{0 dB} & \textbf{-5 dB} & \textbf{-10 dB} \\
        \hline
        MHAtt-RNN  & 784 & 141.6 & 98.04 & 78.83 & 74.25 & 61.98 & 55.87 & 50.98 \\
        KWT-1  & 607 & 53.8 & 97.72 & 88.73 & 85.46 & 73.52 & 67.59 & 59.07 \\
        ResNet-15  & 238 & 961.2 & 96.48 & 89.45 & 87.34 & 79.00 & 73.58 & 66.73 \\
        MatchboxNet-6x2x64  & 140 & 36.8 & 97.60 & 87.34 & 85.19 & 75.58 & 70.06 & 62.35 \\
        BC-ResNet-6  & 202 & 55.4 & 98.20 & 93.67 & 93.30 & 87.48 & 83.06 & 79.40 \\
        ConvMixer   & 119 & 22.2 & 98.20 & 96.32 & 95.52 & 90.62 & 86.37 & 77.47 \\
        FCA-Net (Ours)  & 119 & 22.3 & 98.25 & \textbf{96.96} & \textbf{95.99} & \textbf{92.65} & \textbf{87.97} & \textbf{80.40} \\
        \hline
        AST-Tiny  & 5,805 & 782.2 & 97.65 & 91.02 & 87.71 & 83.31 & 78.95 & 72.32 \\
        KWT-3  & 5,361 & 526.3 & 98.54 & 93.47 & 91.08 & 83.97 & 78.45 & 71.08 \\
        \hline
    \end{tabular}

    \label{tab:comparison}
\end{table*}
\begin{table}[htbp]
\centering
\caption{Exploration of different attention modules and positions of the performance }
\label{tab:my-table}
\resizebox{\columnwidth}{!}{%
\begin{tabular}{c|c|ccccc}
\hline
\multirow{2}{*}{\textbf{Attention}} &
  \multirow{2}{*}{\textbf{Position}} &
 
  \multicolumn{5}{c}{\textbf{Accuracy of Noisy Test Command (\%)}} \\ \cline{3-7} 
   &
   &
  \multicolumn{1}{c|}{\textbf{Clean}} &
  \multicolumn{1}{c|}{\textbf{20 dB}} &
  \multicolumn{1}{c|}{\textbf{0 dB}} &
  \multicolumn{1}{c|}{\textbf{-5 dB}} &
  \textbf{-10 dB} \\ \hline
-                    & -     &   \multicolumn{1}{c|}{96.32} & \multicolumn{1}{c|}{95.52} & \multicolumn{1}{c|}{90.62} & \multicolumn{1}{c|}{86.37} & 77.47 \\ \hline
\multirow{4}{*}{ECA} & pre   &   \multicolumn{1}{c|}{95.23} & \multicolumn{1}{c|}{94.49} & \multicolumn{1}{c|}{89.37} & \multicolumn{1}{c|}{85.24} & 76.97 \\
                     & post    & \multicolumn{1}{c|}{96.77} & \multicolumn{1}{c|}{95.83} & \multicolumn{1}{c|}{90.75} & \multicolumn{1}{c|}{86.20} & 78.12 \\
                     & all     & \multicolumn{1}{c|}{96.73} & \multicolumn{1}{c|}{95.68} & \multicolumn{1}{c|}{91.86} & \multicolumn{1}{c|}{86.60} & 80.00 \\
                     & final   & \multicolumn{1}{c|}{96.67} & \multicolumn{1}{c|}{95.85} & \multicolumn{1}{c|}{91.14} & \multicolumn{1}{c|}{85.90} & 78.27 \\ \hline
\multirow{4}{*}{SE}  & pre     & \multicolumn{1}{c|}{95.23} & \multicolumn{1}{c|}{94.59} & \multicolumn{1}{c|}{89.70} & \multicolumn{1}{c|}{84.50} & 75.16 \\
                     & post  &   \multicolumn{1}{c|}{94.70} & \multicolumn{1}{c|}{94.22} & \multicolumn{1}{c|}{89.64} & \multicolumn{1}{c|}{84.89} & 76.85 \\
                     & all   &  \multicolumn{1}{c|}{95.50} & \multicolumn{1}{c|}{94.86} & \multicolumn{1}{c|}{89.82} & \multicolumn{1}{c|}{85.20} & 76.42 \\
                     & final &   \multicolumn{1}{c|}{95.20} & \multicolumn{1}{c|}{94.10} & \multicolumn{1}{c|}{89.97} & \multicolumn{1}{c|}{85.51} & 76.75 \\ \hline
\multirow{4}{*}{C2D} & pre   &   \multicolumn{1}{c|}{95.70} & \multicolumn{1}{c|}{94.47} & \multicolumn{1}{c|}{89.08} & \multicolumn{1}{c|}{84.07} & 75.88 \\
                     & post  &   \multicolumn{1}{c|}{95.27} & \multicolumn{1}{c|}{94.65} & \multicolumn{1}{c|}{89.02} & \multicolumn{1}{c|}{84.00} & 75.56 \\
                     & all   &   \multicolumn{1}{c|}{96.96} & \multicolumn{1}{c|}{95.99} & \multicolumn{1}{c|}{92.65} & \multicolumn{1}{c|}{87.97} & 80.40 \\
                     & final &   \multicolumn{1}{c|}{96.92} & \multicolumn{1}{c|}{96.09} & \multicolumn{1}{c|}{91.18} & \multicolumn{1}{c|}{86.82} & 78.42 \\ \hline
\end{tabular}%
}
\end{table}
\section{Experiment setting}
\subsection{Dataset for Noisy Keyword Spotting}
We evaluate our proposed system on the Google Speech Commands V2 dataset, which contains 105,000 utterances of 35 unique words, each 1 second long and sampled at 16 kHz. We use the official train, validation, and test split for the 12-label classification task. These labels include: `up', `down', `left', `right', `yes', `no', `on', `off', `go', `stop', `silence', and `unknown'. The `unknown' class includes the remaining words in the dataset.

To simulate a noisy far-field environment, we use noise samples from the MUSAN dataset, which has 930 files of assorted noises, totaling about 6 hours, all sampled at 16 kHz. These noises include technical and non-technical sounds like DTMF tones, thunder, and car horns. We add these noises to our command samples to mimic various noisy conditions. This process is divided into four progressively harder steps. Initially, the model is trained on clean samples without noise. In the following three steps, noise is incrementally introduced to the fixed $N$ samples in decrements of -5dB, with conditions in $N$ samples uniformly distributed, i.e., [clean, 0], [clean, 0, -5], [clean, 0, -5, -10].

\subsection{Implement details}
We use Mel-Frequency Cepstral Coefficients (MFCC) with the following parameters: sample rate of 16 kHz, 40 MFCCs, and mel-spectrogram settings of n\_fft=400, hop\_length=160, n\_mels=64, f\_min=20, and f\_max=8000. Each command is 1 second long; shorter commands are zero-padded to the right. During training, we perform data augmentation with time shifts of -100 to 100 ms and spectrogram masking with maximum lengths of 25 for both time and frequency. We generate noisy data with SNR of 0, -5, and -10 dB.

For stronger regularization, we use input mixup with a ratio of 0.2 on training samples. The model is trained with a batch size of 128 and an initial learning rate of 0.005, reduced by a factor of 0.85 every four epochs after the fifth epoch. We use the Adam optimizer and a learning rate scheduler with Cosine Annealing Warm Restarts. The model is trained for 200 epochs with early stopping if there is no improvement for 20 epochs. This study applied three augmentation methods: mixup~\cite{zhang2018mixup}, SpecAugment~\cite{park2019specaugment}, and curriculum learning~\cite{braun2017curriculum, ng2022convmixer}.

\subsection{Baselines}
We also set up some SOTA keyword spotting models as the baseline to analyze our proposed model. Here are the models we used for experiments:

\begin{itemize}
    \item \textbf{MHAtt-RNN:} This model integrates multi-head attention mechanisms with recurrent neural networks to enhance sequence learning and capture long-range dependencies.
    \item \textbf{KWT:} Keyword Transformer (KWT-1) uses transformer architecture to identify keywords, leveraging self-attention for efficient feature extraction. KWT-3  improves upon KWT-1 by incorporating more layers and sophisticated attention mechanisms for enhanced keyword detection.
    \item \textbf{AST-Tiny:} Audio Spectrogram Transformer-Tiny (AST-Tiny) employs a transformer architecture specifically designed for processing audio spectrograms, focusing on smaller model sizes.
    \item \textbf{BC-ResNet-6:} Broadcasted Residual Learning Network (BC-ResNet-6) incorporates broadcasted residual connections over 6 layers to enhance the learning of spatial features.
    \item \textbf{ResNet-15:} A variant of the ResNet architecture with 15 layers, ResNet-15 employs residual connections to facilitate deeper networks and improve feature learning.

    \item \textbf{MatchboxNet-6\(\times\)2\(\times\)64:} This model uses a combination of 6 blocks, each containing 2 convolutional layers with 64 filters, optimized for efficient keyword spotting.
    
    \item \textbf{ConvMixer:} ConvMixer combines convolution operations with mixer layers to extract and mix features across both frequency and temporal dimensions.
\end{itemize}
\section{Results}

\subsection{Compare study among the SOTA models}

We compare the performance of FCA-Net with previously proposed SOTA models. Models are retrained from the official source code provided with our designed data environment. The results are shown in Table II. From the table, we observe that our proposed model achieved the highest accuracy among small models when tested on the official V2-12, with an accuracy improvement of approximately 2\% to 3\% over other small models like ConvMixer. Furthermore, FCA-Net shows a noticeable reduction in the number of model parameters and MACs, indicating lower memory and computation resource requirements. Most importantly, when evaluated in noisy conditions, FCA-Net demonstrates superior performance, achieving up to 7.4\% better accuracy compared to other models with similar memory footprints. The proposed model is also competitive with larger transformer-based models (KWT-3, AST-Tiny) under challenging noisy conditions, showcasing its efficiency and reliability for real-world applications.

\subsection{Ablation study}
The results indicate that incorporating efficient attention modules significantly enhances the robustness of KWS systems in noisy environments. Among the different positions tested, the C2D attention module demonstrates superior performance across various noise levels. Notably, the C2D module achieves the highest accuracy when placed at all positions within the ConvMixer block, consistently outperforming the other attention methods, including ECA and SE. This setup proves to be the most effective strategy for enhancing noise robustness, making the C2D module particularly valuable for improving KWS system performance in challenging auditory conditions.

\section{Conclusions}
In this paper, we proposed FCA-Net, a novel convolutional neural network designed to enhance the robustness of Keyword Spotting (KWS) systems in noisy environments while maintaining a small memory footprint. Our approach combines mixer unit-based feature interaction with an efficient two-dimensional convolution-based attention module, further strengthened by a curriculum-based multi-condition training strategy. Experimental results demonstrate that FCA-Net achieves state-of-the-art accuracy among small models and outperforms existing solutions in noisy conditions. Additionally, FCA-Net significantly reduces the number of parameters and computational resources required, making it both efficient and effective for real-world applications. Our findings highlight the potential of FCA-Net as a reliable and resource-efficient solution for robust KWS in diverse and challenging auditory environments.










\clearpage
\printbibliography
\end{document}